\documentclass[english,twocolumn,prl,superscriptaddress,longbibliography]{revtex4-1}
\usepackage[T1]{fontenc}
\usepackage[utf8]{inputenc}
\usepackage{color}
\usepackage{amsmath}
\usepackage{graphicx}
\usepackage{bbm}

\usepackage[colorlinks=true,linkcolor=blue,urlcolor=blue,citecolor=blue,anchorcolor=blue]{hyperref}
\makeatletter
\usepackage{amsmath}
\usepackage{subfigure}
\usepackage{graphicx, mathtools}
\usepackage[normalem]{ulem}
\usepackage{soul}

\graphicspath{ {Figures/} }

\makeatother
\usepackage{amsfonts}
\usepackage{babel}
\begin{document}
\newcommand{\be}{\begin{equation}}
\newcommand{\ee}{\end{equation}}
\newcommand{\bn}{\begin{eqnarray}}
\newcommand{\en}{\end{eqnarray}}
\newcommand{\ii}{\'{\i}}
\newcommand{\ca}{\c c\~a}
\newcommand{\uc}{\uppercase}
\newcommand{\tb}{\textbf}
\newcommand{\bw}{\begin{widetext}}
\newcommand{\ew}{\end{widetext}}
\newcommand{\blue}{\textcolor{blue}}
\newcommand{\red}{\textcolor{red}}
\newcommand{\magenta}{\textcolor{magenta}}
\newcommand{\new}[1]{\textcolor{red}{#1}}
\newcommand{\ChM}[1]{\textcolor{blue}{#1}}
\newcommand{\comm}[1]{\textcolor{green}{(GL: #1.)}}
\newcommand{\dbra}{\big\langle\!\!\big\langle}
\newcommand{\dket}{\big\rangle\!\!\big\rangle}

\title{Prethermalization and wave condensation in a nonlinear disordered Floquet system}

\author{Prosenjit Haldar}\email{prsnjthldr@gmail.com}
\affiliation{Laboratoire de Physique Th\'{e}orique, Universit\'{e} de Toulouse, CNRS, UPS, France}

\author{Sen Mu}\email{senmu@u.nus.edu}
\affiliation{Department of Physics, National University of Singapore, 2 Science Drive 3, Singapore 117542, Singapore}

\author{Bertrand Georgeot}
\affiliation{Laboratoire de Physique Th\'{e}orique, Universit\'{e} de Toulouse, CNRS, UPS, France}

\author{Jiangbin Gong}
\affiliation{Department of Physics, National University of Singapore, 2 Science Drive 3, Singapore 117542, Singapore}

\author{Christian Miniatura}
\affiliation{MajuLab, International Joint Research Unit IRL 3654, CNRS, Universit\'e C\^ote d'Azur, Sorbonne Universit\'e, National University of Singapore, Nanyang
Technological University, Singapore}
\affiliation{Centre for Quantum Technologies, National University of Singapore, 117543 Singapore, Singapore}
\affiliation{Department of Physics, National University of Singapore, 2 Science Drive 3, Singapore 117542, Singapore}
\affiliation{School of Physical and Mathematical Sciences, Nanyang Technological University, 637371 Singapore, Singapore}
\affiliation{Université Côte d’Azur, CNRS, INPHYNI, Nice, France}

\author{Gabriel Lemari\'e}\email{lemarie@irsamc.ups-tlse.fr}
\affiliation{MajuLab, International Joint Research Unit IRL 3654, CNRS, Universit\'e C\^ote d'Azur, Sorbonne Universit\'e, National University of Singapore, Nanyang
	Technological University, Singapore}
\affiliation{Centre for Quantum Technologies, National University of Singapore, 117543 Singapore, Singapore}
\affiliation{Laboratoire de Physique Th\'{e}orique, Universit\'{e} de Toulouse, CNRS, UPS, France}

\begin{abstract}
		Periodically-driven quantum systems make it possible to reach stationary states with new emerging properties. However, this process is notoriously difficult in the presence of interactions because continuous energy exchanges generally boil the system to an infinite temperature featureless state. Here, we describe how to reach nontrivial states in a periodically-kicked Gross-Pitaevskii disordered system. One ingredient is crucial: {both} disorder and {kick strengths} should be weak enough to induce sufficiently narrow and well-separated Floquet bands. In this case, inter-band heating processes are strongly suppressed and the system can reach an exponentially long-lived prethermal plateau described by the Rayleigh-Jeans distribution. Saliently, {the system can even undergo a wave condensation process when its initial state has a sufficiently low total quasi-energy}. These predictions {could be tested in nonlinear optical experiments or with ultracold atoms}.
\end{abstract}

\maketitle

{\it Introduction.--}  
The nonequilibrium dynamics of periodically driven quantum systems has been under intensive scrutiny over the past few years~\cite{Floquet1, oka2019floquet, harper2019topology, bukov2015universal, moessner2017equilibration}. It has been shown that interesting stationary states could emerge after appropriate Floquet engineering~\cite{Topology_solid_state2, Topology_theory1,Topology_theory6, Topology_solid_state1, Topology_solid_state3, Topology_cold_atoms1, Topology_cold_atoms2,Topology_cold_atoms3, Topology_cold_atoms4, Topology_photonic1, Topology_photonic2,PhysRevLett.109.010601,mciver2020light,wintersperger2020realization}. However, the interplay between interactions and temporal driving generally induces heating processes which force the system into a featureless state of infinite temperature~\cite{many_body_heating_experiment,many_body_heating2, PONTE2015196}. 
Different strategies have been discussed to prevent this detrimental heating, e.g. many-body localization~\cite{Floquet_MBL2,Floquet_MBL3,Floquet_MBL1,Sondhi_PRL, time_crystals, PhysRevB.94.085112}, or coupling to a bath~\cite{Bath_coupling1,Bath_coupling2}. An alternate route is {to use high-frequency driving to maintain the system in a long-lived metastable state before heating to infinite temperature takes over.}  Known as prethermalization, this strategy works for both quantum ~\cite{Bukov2015,Abanin2015,Mori2016,Else2017,Mallayya2019,Bloch2020,Fleckenstein2021,Peng2021} and classical systems~\cite{Mori2018,Rajak2019,Howell2019,Hodson2021}. 

In this Letter, we consider a variant of the quantum kicked rotor (QKR), a paradigmatic model of quantum chaos \cite{quantum_chaos, IZRAILEV1990299}, in the presence of a weak nonlinear Gross-Pitaevskii (GP) interaction term. In the absence of interactions, the QKR
shows dynamical localization \cite{casati1979stochastic}, a phenomenon analogous to Anderson localization \cite{50Years} but in momentum ($p$) space \cite{Fishman} {which has been observed in a number of cold atom experiments~\cite{moore1994atom, moore1995atom, Chabe2008MIT, QQKR, GARREAU201731}. The presence of GP interaction terms challenges dynamical localization: transport is no longer frozen and anomalous diffusion rules the spreading of wavepackets \cite{QKR_GPE1,QKR_GPE2,Cherroret2014,PhysRevA.101.043624}. Analogous observations have been made in disordered nonlinear Schr\"odinger chains, see e.g. \cite{PhysRevLett.100.094101, PhysRevE.79.026205, PhysRevLett.111.064101, PhysRevE.86.036202, Cherroret_2016, PhysRevLett.124.186401}}.

On the other hand, it is known that the propagation of a random initial wave in a nonlinear medium can give rise to Rayleigh-Jeans (RJ) thermalization and wave condensation phenomena as observed in nonlinear optics~\cite{Thermal_BEC1,Thermal_BEC2,Aschieri2011} and explained by wave turbulence theory \cite{WT_theory1,WT_theory2, Naza2015}{. 
A similar situation happens when an initial plane wave propagates in a disordered nonlinear medium, disorder inducing the required wave randomization \cite{Thermal_BEC3}. This configuration is particularly interesting because (i) it is the natural setting to observe the coherent backscattering (CBS) \cite{CBS1, CBS2, CBS3, CBS4} and coherent forward scattering (CFS) effects \cite{CFS1, CFS2, CFS3, CFS4, CBS_CFS_G, PhysRevB.97.041406, martinez2020coherent} when interactions are absent and (ii) it allows to control the initial energy of the wave and thus the route to wave condensation in nonlinear systems.}

In this Letter, we study RJ thermalization and wave condensation phenomena for the nonlinear QKR and establish when they happen in this Floquet system: the quasi-energies of the system should form sufficiently narrow bands as compared to their separation. This {happens} when the kick and disorder strengths are smaller than a characteristic threshold. In this case, heating processes associated to inter-band transitions are strongly suppressed and {the system can reach} an exponentially long-lived prethermal plateau where RJ thermalization takes place. Moreover, wave condensation at the bottom of the fundamental Floquet band is observed. Ultimately, at exponentially large times, {the system heats up to the infinite-temperature state}.

\begin{figure}
\includegraphics[width=1.0\linewidth]{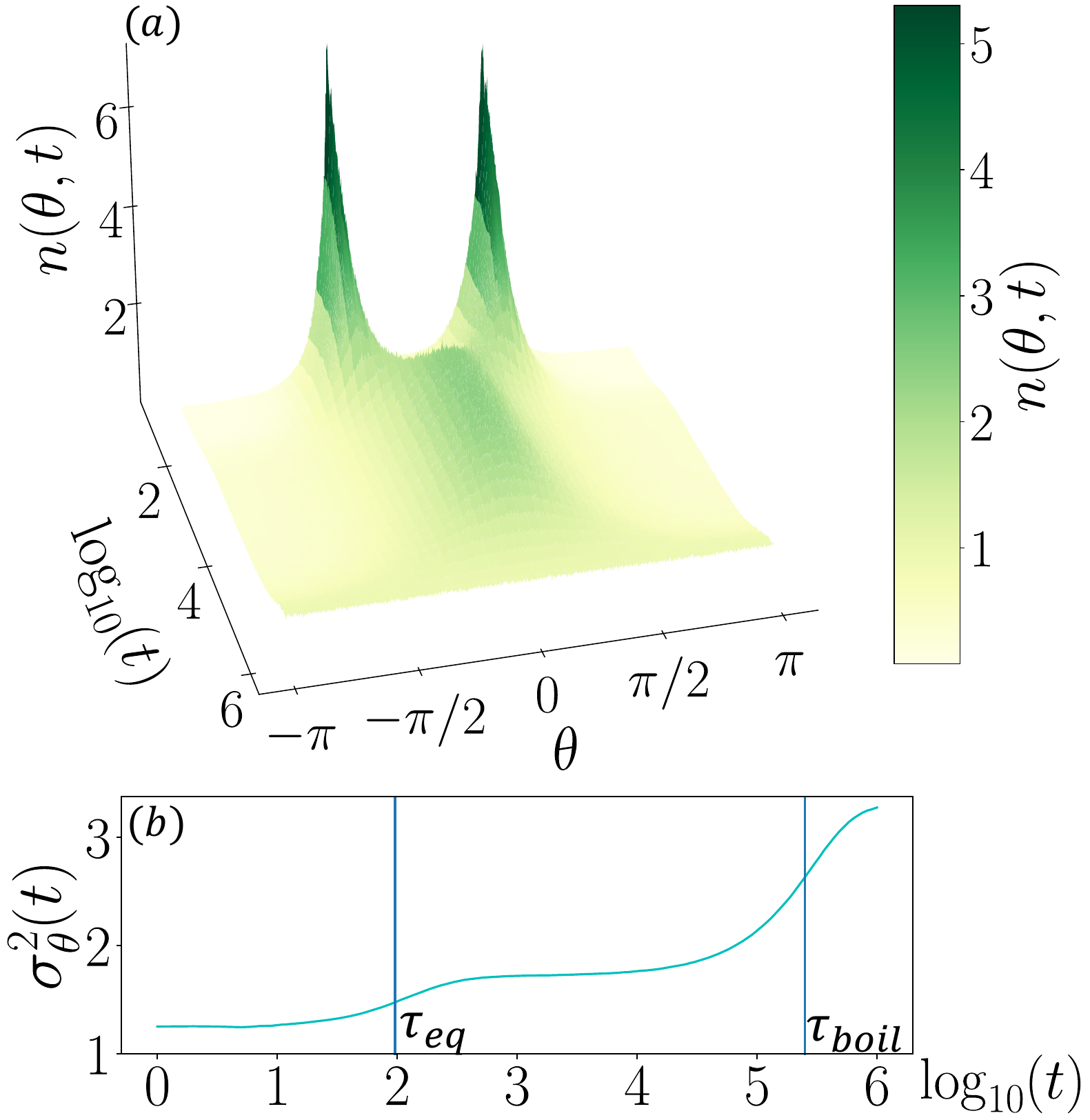}
\caption{Nonequilibrium dynamics of the nonlinear disordered Floquet system \eqref{eq:KRGP}. (a) $\theta$-distribution $n(\theta,t)$ obtained when interaction is switched on after dynamical localization, signalled by the CBS-CFS peaks at $\pm \theta_0$, is achieved (see text). Nonlinearity quickly erases these CBS-CFS interference effects. For sufficiently small $W$ and $K$, $W+K\lesssim \pi/2$, the system reaches a prethermal equilibrium  when $t > \tau_{eq}$, with a quasi-stationnary $n_{eq}(\theta)$. Ultimately, after a time $\tau_{boil}$ which can be exponentially large, Floquet heating brings the system to an infinite temperature state with uniform $n_\infty(\theta)$. (b) The variance $\sigma^2_\theta(t)$ of $n(\theta,t)$ clearly shows the 3 dynamical regimes. The parameters are $K=1.0$, $W=0.4$, $g=0.1$, $\theta_0=1.05$, $N=1024$ and $N_d=1200$.
}
\label{fig:time_evol}
\end{figure}

{\it Model.--} Our model {features} the nonlinear QKR with a GP interaction term in $p$-space \cite{QKR_GPE1}:
\begin{equation}\label{eq:KRGP}
i \partial_t \psi(p,t)= \mathcal{H}(t) \psi(p,t) + g {N_a} \vert \psi(p,t) \vert^2 \psi(p,t)
\end{equation}
where $\mathcal{H}(t) = \hat{p}^2/2 - K \cos \theta \sum_{j} \delta(t-j)$ ($\hbar=1$, moment of inertia $I=1$).
Here, $\theta \in [-\pi, \pi)$ is the rotor angle, $\hat{p}=-i\partial_\theta$ the momentum, $K$ the kick strength, $g$ the nonlinear strength{, and $N_a$ the number of atoms}. 
Because 
{$\psi(\theta + 2\pi,t) = \psi(\theta,t)$}, we have $\psi(\theta,t) = \sum_{p} \psi(p,t) \, e^{\mathrm{i} p\theta}$ {where momentum states are labeled by integers $p\in\mathbbm{Z}$}. {Our normalization reads $\int_{-\pi}^{\pi}|\psi(\theta,t)|^2 d\theta/(2\pi) = \sum_p |\psi(p,t)|^2=1$. In our numerical simulations, we have considered a finite-size momentum basis set to $|p| \leq N/2$ with periodic boundary conditions in $p$-space, and an atomic density equal to unity, i.e. $N_a=N$.}

Defining $\psi^{\pm}_{n}(\cdot) = \psi(\cdot, t\!=\!n\!\pm\!0^{+})$, the Floquet dynamics associated to Eq.~\eqref{eq:KRGP} is obtained by iterating the nonlinear map
\begin{eqnarray}
\psi^{-}_{n+1} (p) &=& e^{-\mathrm{i} \alpha(p)} \, e^{-\mathrm{i} g N \vert \psi^{+}_{n}(p)\vert^2} \, \psi^{+}_{n} (p), \\
\psi^{+}_{n+1} (\theta) &=& e^{\mathrm{i} K \cos\theta} \, \psi^{-}_{n+1}(\theta).
\label{Floquet}
\end{eqnarray} 
Here, we have replaced the quasi-random phases induced by the kinetic term $\hat{p}^2/2$ by true random phases $\alpha (p)$. Usually, they are taken from a uniform distribution within $[-\pi, \pi]$ \cite{Fishman,George_B}. In our study, we crucially consider smaller intervals $[-W, W]$, where $W \leq \pi$ is the disorder strength.
{The motivation for choosing local interactions in $ p $ is now clear: First, it transposes to $p$-space the usual case where disorder and interactions are local in position (see e.g.~\cite{Thermal_BEC3}). Second, the associated quantum map Eq.\eqref{Floquet} allows to evolve systems of large sizes $N \sim 10^5$ up to large times $t \sim 10^6$ at low numerical cost.}

Without interaction, {localization occurs in $ p $-space and CBS and CFS take place} in $ \theta $-space~\cite{CBS_CFS_G}. Starting from an initial rotor angle $\theta_0$, {$\psi^{+}_0(p)=\frac{1}{\sqrt{N}}\exp(-\mathrm{i}p\theta_0)$}, a CBS peak appears at $-\theta_0$ on top of a diffusive background over the Boltzmann transport time $\tau_B$. The CFS peak emerges at $\theta_0$ when dynamical localization sets in, i.e. after the Heisenberg time $\tau_H$. Denoting disorder average by $\overline{(\cdot\cdot\cdot)}$ {and the number of disorder configurations considered by $N_d$}, the angle distribution $n(\theta,t)={\overline{\vert \psi(\theta,t)\vert^2}}$ becomes stationary at times $t\gg \tau_H$ and consists of twin CBS and CFS peaks over {a diffusive} background.

Following the analogy with thermalization and condensation predicted in disordered nonlinear systems~\cite{Thermal_BEC3}, we consider the effect of interactions {on these peaks}. Hereafter, we set the GP interaction and disorder strengths to $g=0.1$ and $W=0.4$ respectively and we study the role of the kick strength $K$ and initial rotor angle $\theta_0$ on the dynamics of the system. Depending on $K$ and $\theta_0$, we find 3 different dynamical behaviors: (i) the system quickly adopts a uniform $\theta$-distribution $n_\infty(\theta) = 1$ ("infinite temperature" featureless state); (ii) the system transits through a long-lived metastable state characterized by a nontrivial $\theta$-distribution $n_{eq}(\theta)$ centered at $\theta=0$ and related to the thermal RJ distribution~\cite{Baudin2020} before eventually going to $n_\infty(\theta)$;
and (iii) the system enters a condensation regime and develops a peak structure at $\theta=0$ on top of $n_{eq}(\theta)$ before, again, eventually going to $n_\infty(\theta)$. Crucially, the life time of the metastable state can be made exponentially large, allowing for the observation of thermalization or condensation. 

{\it Characteristic time scales.--}
{The different characteristic time scales are illustrated in Fig.~\ref{fig:time_evol}.
First,} the CBS and CFS peaks quickly decay over a characteristic time $\tau_{g} \propto g^{-1}$ (see Supplementary Material (SM) and \cite{Cherroret2014,PhysRevResearch.2.033349}), signaling that the GP interaction term is indeed wiping out disorder-induced interference effects. Meanwhile, the GP term redistributes the energy over the different Floquet modes and the system reaches a so-called prethermalization plateau after some equilibration time $\tau_{eq}\sim g^{-2}$ (see SM and \cite{Thermal_BEC3}) where it stabilizes in a metastable state with {distribution} $n_{eq}(\theta)$. The system gets eventually boiled, around time $\tau_{boil}$, to the infinite temperature state $n_\infty(\theta)=1$. The lower panel of Fig.~\ref{fig:time_evol} clearly shows the equilibration, metastable and heating regimes by plotting the variance $\sigma^2_\theta (t) =\langle \theta^2\rangle -\langle \theta \rangle^2$ of $n(\theta,t)$ as a function of time ($\sigma^2_\theta(t\rightarrow \infty) = \pi^2/3$ for $n_\infty(\theta)=1$).  
Hereafter, we focus on the parameter sector $\tau_g \ll \tau_H$ and $\tau_\text{eq} \gg \tau_B, \tau_g$. In this case, interference effects are negligible and {multiple scattering randomizes the wave much faster than the system equilibrates.}

\begin{figure}
	\includegraphics[width=\linewidth]{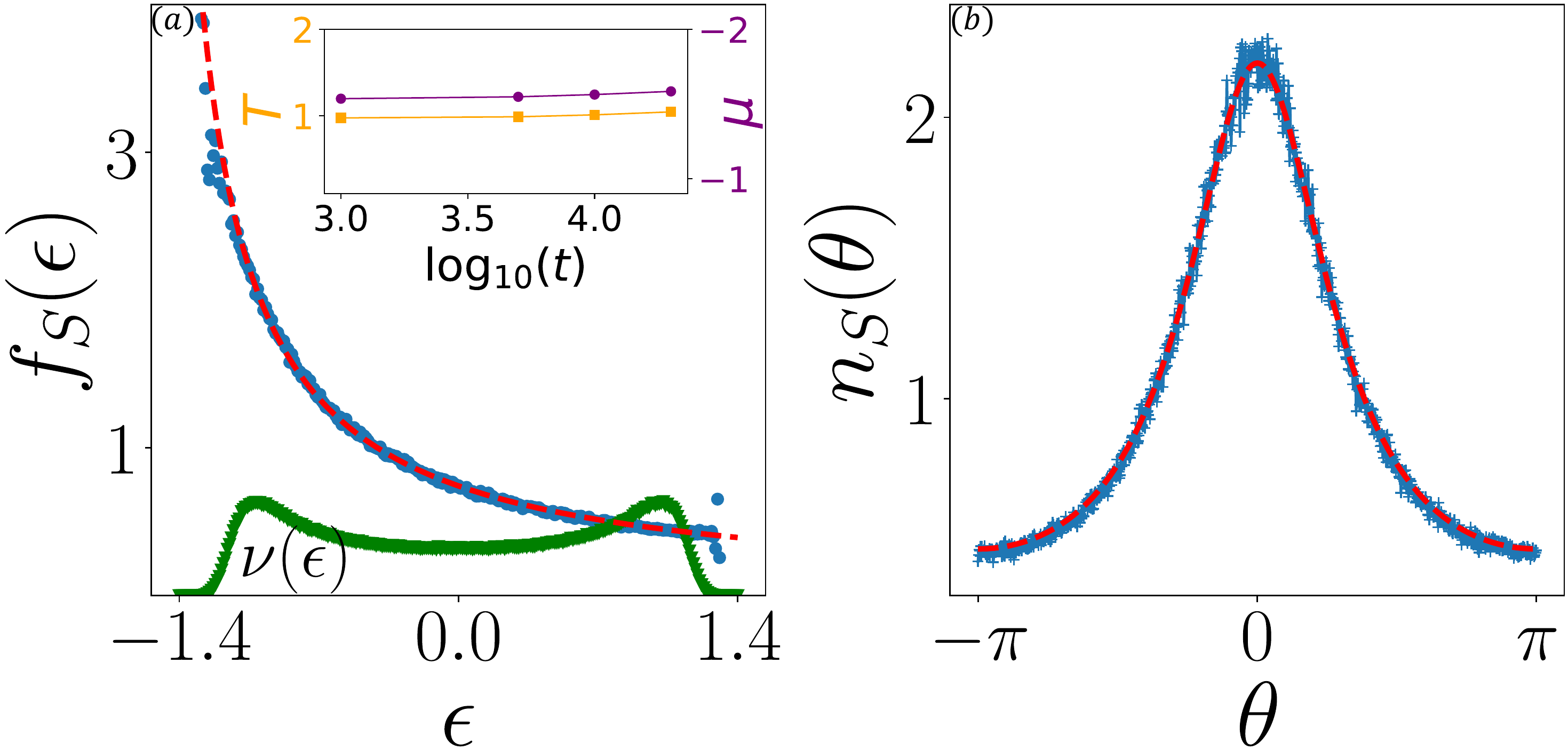}
	\caption{Prethermal state obtained at $t=10^4$ for large $\theta_0=1.05$ and $K+W <\pi/2$. (a) Quasi-energy distribution $f_S(\varepsilon)$ (upper blue dots) and density of states $\nu(\varepsilon)$ (lower green dots). The Rayleigh-Jeans distribution $f_{RJ}(\varepsilon)$ with $T= 1.17$ and $\mu=-1.58$ (red dashed line) fits very well the data. Inset: time dependence of $T$ and $\mu$. (b) $\theta$-distribution $n_S(\theta)$ (blue dotted curve) and $n_{eq}(\theta)$ Eq.~\eqref{eq:xdist} (red dashed curve) corresponding to $f_{RJ}(\varepsilon)$. 
			The parameters are $K=1.0$, $W=0.4$, $g=0.1$, $N=1024$ and $N_d=1200$. 
	}
	\label{fig:thermal}
\end{figure}

{\it Floquet-Boltzmann {kinetic equation}.--} 
The non-interacting system is characterized by the Floquet Hamiltonian $\mathcal{H_F}\equiv \mathcal{H}(t) - i\partial_t$. Its linear modes $\vert \phi_\alpha (t)\rangle$ solve $\mathcal{H_F} \vert \phi_\alpha (t)\rangle = \varepsilon_\alpha \vert \phi_\alpha (t)\rangle$, $\varepsilon_\alpha$ being called a quasi-energy. $\mathcal{H}(t)$ and $\vert \phi_\alpha (t)\rangle$ have the same time-periodicity and $\varepsilon_\alpha$ is defined modulo the driving frequency $\Omega=2\pi$ here. The support of the quasi-energy spectrum is $\varepsilon_{\alpha} \in [-E_D,E_D]$ with $E_D= K + W$ for {$E_D\le \pi$.} 

Expanding the state of the interacting system over the Floquet modes, $\vert \psi(t)\rangle = \frac{1}{\sqrt{N}} \sum_{\alpha=1}^N c_\alpha(t) \vert  \phi_\alpha(t)\rangle$, {the term $g N \vert \psi(p)\vert^2$} in Eq.~\eqref{Floquet} redistributes populations among the Floquet modes. The central quantity is therefore the quasi-energy distribution $f_{\varepsilon}(t) \equiv \frac{1}{N} \overline{\sum_{\alpha=1}^{N} \delta(\varepsilon - \varepsilon_\alpha) \vert c_\alpha(t)\vert^2} / \nu(\varepsilon)$, where $\nu(\varepsilon) = \frac{1}{N} \overline{\sum_\alpha \delta(\varepsilon-\varepsilon_\alpha)}$ is the density of states of the linear kicked rotor. It satisfies $\int_{-\pi}^{\pi} d\varepsilon \, \nu(\varepsilon) f_{\varepsilon}(t) =1$ since $\sum_{\alpha} \vert c_\alpha(t)\vert^2 = N$. In the limit $\tau_g \ll \tau_H$ and $\tau_\text{eq} \gg \tau_B$, the $\theta$-distribution at time $t \in \mathbb{N}$ reads \cite{Thermal_BEC3}:
\begin{equation}
  n(\theta,t) \approx \int_{-\pi}^{\pi} d\varepsilon \, A_\varepsilon(\theta) f_{\varepsilon}(t),  
  \label{eq:xdist}
\end{equation}
{where $A_\varepsilon(\theta)=\frac{1}{N} \overline{\sum_{\alpha} \delta(\varepsilon-\varepsilon_{\alpha}) \vert \phi_{\alpha}(\theta,0)\vert^{2}}$ is the spectral function.}

{For time-independent systems, a kinetic equation {for $f_{\varepsilon}(t)$} can be derived under certain approximations \cite{WT_theory2, Naza2015, PhysRevLett.124.186401},} 
\begin{align}
\partial_t f_{\varepsilon} =  & \, 4\pi g^2 \int  d\varepsilon_2d\varepsilon_3d\varepsilon_4 \ R(\varepsilon,\varepsilon_2,\varepsilon_3,\varepsilon_4) \  \nu(\varepsilon_2)\nu(\varepsilon_3)\nu(\varepsilon_4) \nonumber\\
& [f_{\varepsilon}f_{\varepsilon_3}f_{\varepsilon_4} + f_{\varepsilon_2}f_{\varepsilon_3}f_{\varepsilon_4} - f_{\varepsilon}f_{\varepsilon_2}f_{\varepsilon_3} - f_{\varepsilon}f_{\varepsilon_2}f_{\varepsilon_4}].
\label{eq:kinetic}
\end{align}
The 4-wave mixing collision kernel $R$ imposes energy conservation for each energy-exchange process ($\varepsilon+\varepsilon_2 = \varepsilon_3+\varepsilon_4$). We {can} generalize this description to Floquet systems by expanding the time-periodic Floquet modes in Fourier series $\phi_\alpha^{(m)}(p) = \int_0^1 dt \, \phi_\alpha(p,t) \exp(-2i\pi m t)$ \cite{many_body_heating8,slow_heating2}, thereby incorporating the so-called Umklapp processes \cite{kittel1976introduction} describing inter-band transitions:
\begin{eqnarray}
R = \sum_{m \in \mathbbm{Z}} R^{(m)}(\varepsilon,\varepsilon_2,\varepsilon_3,\varepsilon_4) \delta(\varepsilon\!+\!\varepsilon_2\!-\!\varepsilon_3\!-\!\varepsilon_4\!+\!2\pi m),
 \label{eq:collisionK}
\end{eqnarray}
see SM and \cite{WT_theory3,Thermal_BEC3}. The kernels $R^{(m)}$
relate to an overlap between four Fourier Floquet mode amplitudes. 

\begin{figure}
	\includegraphics[width=\linewidth]{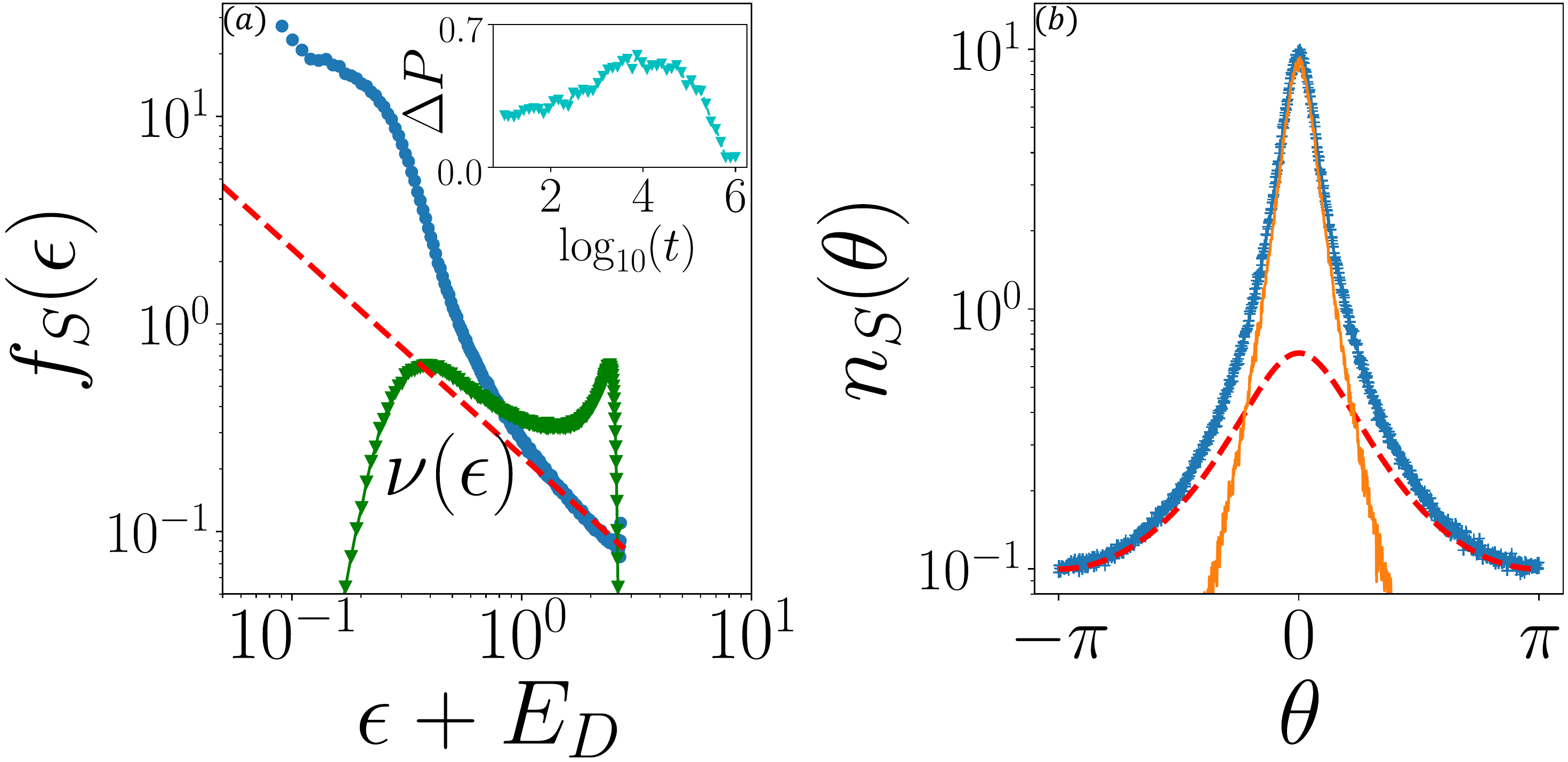}
	\caption{Wave condensed state obtained at $t=10^4$ for small $\theta_0=0.32$ and $K+W<\pi/2$. (a) {$f_{S}(\varepsilon)$ (upper blue dots) is compared to the best fit $f_{RJ}(\varepsilon)$ (red dashed line) obtained at temperature $T=0.23$ and computed at the maximal value $\mu = -E_D$. $f_{RJ}(\varepsilon)$ fails to reproduce the data at low $\varepsilon$.} Lower green dots represent $\nu(\varepsilon)$. Inset: time dependence of the population gap $\Delta P$ of the 1-body density matrix. (b) $n_{S}(\theta)$ (blue dotted curve) is the sum of the Rayleigh-Jeans component $n_{eq}(\theta)$ (red dashed curve) and a condensed component $n_{c}(\theta)$ (orange curve). {Its bimodal structure is typical }of condensation (see text). The parameters are $K=1.0$, $W=0.4$, $g=0.1$, $N=1024$ and $N_d=1200$.
	}
	\label{fig:cond}
\end{figure}

{\it {Prethermalization}.--} 
Following this approximate approach, the collision kernel \eqref{eq:collisionK} predicts a suppression of heating when Umklapp processes $m\ne 0$ are forbidden. This condition is met when the Floquet bandwidth $2E_D$ is sufficiently narrow, $E_D=K+W<\pi/2$. 
In this case, we effectively recover the kinetic equation \eqref{eq:kinetic} of time-independent systems
and the total linear contribution of the quasi-energy per mode of the system
$E_{tot}=\int_{-\pi}^{\pi} d\varepsilon \, \varepsilon \, \nu(\varepsilon) \, f_\varepsilon(t)$, see \cite{Thermal_BEC2}, is conserved. 
{An} equilibrium distribution is {then} obtained by canceling the bracketed term in Eq.~\eqref{eq:kinetic}. {One finds} the RJ distribution $f_{RJ}(\varepsilon)=\frac{T}{\varepsilon-\mu}$ \cite{WT_theory2, WT_theory3, Thermal_BEC3} where $T$ and $\mu$ play respectively the role of temperature and chemical potential. {Note that} $\mu\leq - E_D$ since $f_{RJ}(\varepsilon)>0$. 

\begin{figure}
\includegraphics[width=\linewidth]{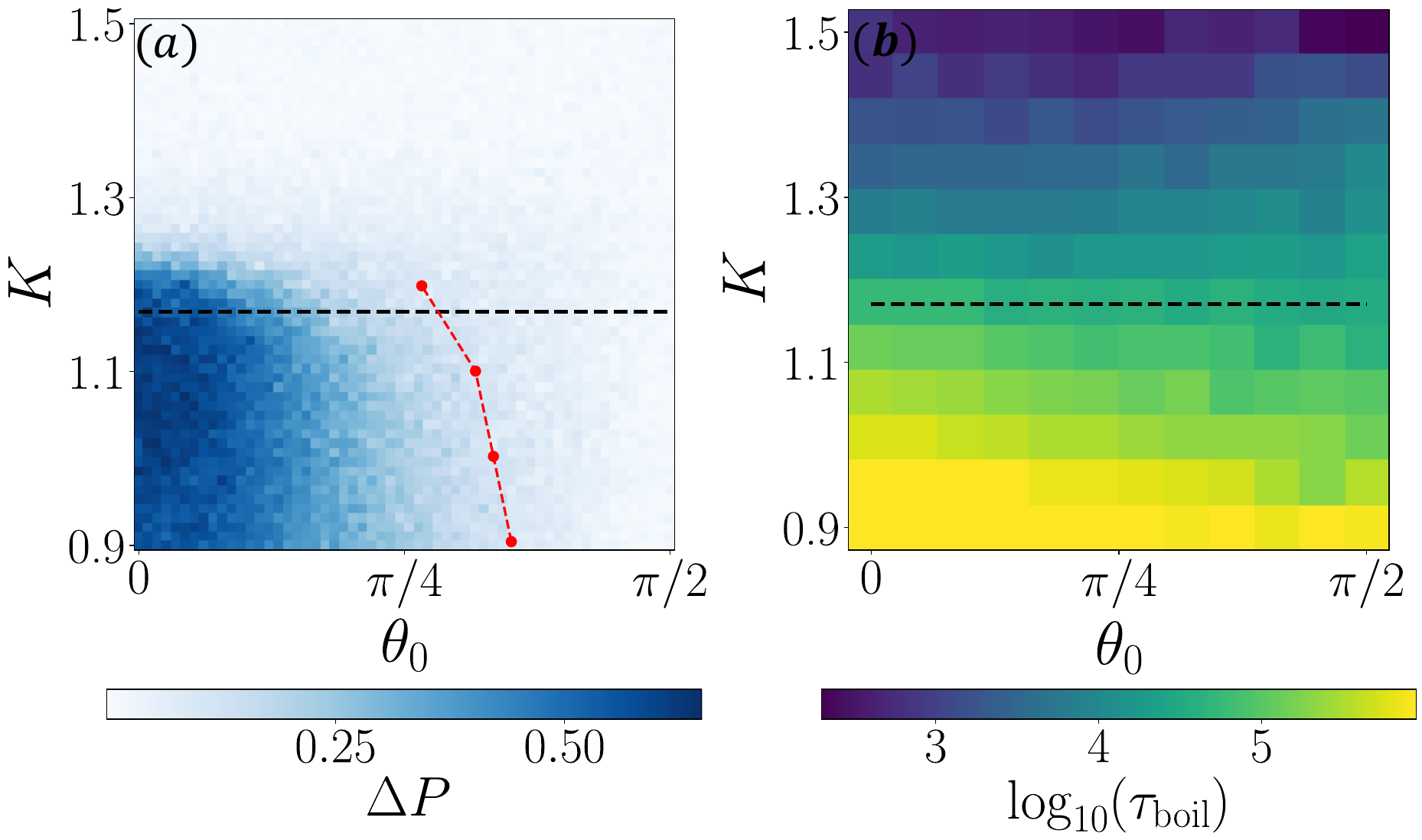}
\caption{Phase diagram of the system in the ($K,\theta_0$) plane. (a) Population gap $\Delta P$ of the $1$-body density matrix at $t=10^4$. The black dashed line represents the heating threshold $E_D=K+W = \pi/2$ of Eq.~\eqref{eq:kinetic}-\eqref{eq:collisionK}. The red dots show the predicted onset $\theta_0^c(K)$ for wave condensation (the line is a guide to the eye). (b) Heating time scale $\tau_{boil}$ after which the infinite-temperature state is reached. The parameters are $W=0.4$ and $g=0.1$, $N=1024$ and $N_d=400$. 
}
\label{fig:phaseportrait}
\end{figure}

Fig.~\ref{fig:thermal}(a) shows the disorder-averaged quasi-energy distribution $f_S(\varepsilon)$ obtained numerically by iterating Eq.~\eqref{Floquet} for $t=10^4$ periods for the parameters used in Fig.~\ref{fig:time_evol} ($E_D=1.4<\pi/2$). As one can see, $f_S(\varepsilon)$ is very well fitted by $f_{RJ}(\varepsilon)$ with $\mu=-1.58$ and $T=1.17$. The corresponding $n_S(\theta)$ is shown in Fig.~\ref{fig:thermal}(b). It agrees very well with the predicted $n_{eq}(\theta)$ obtained from $f_{RJ}(\varepsilon)$ using Eq.~\eqref{eq:xdist}. Moreover, both $T$ and $\mu$, shown in the inset of Fig.~\ref{fig:thermal}(a), stay essentially constant for long times $ \tau_{eq}\ll t \ll \tau_{boil}$. These observations confirm the Floquet-Boltzman approach. However, at times $t \gtrsim \tau_{boil}$, we observe heating: $f_\varepsilon(t)$, $n(\theta,t)$, $E_{tot}$ and $\sigma^2_\theta$ vary rapidly with time (see Figs.~\ref{fig:time_evol},\ref{fig:phaseportrait} and SM). Note that $f_\varepsilon(t)$ can still be fitted by a RJ distribution, with time-dependent $T(t) \rightarrow +\infty$, $\mu(t)\rightarrow -\infty$ and $T(t)/\mu(t) \rightarrow -1$ when $t \rightarrow \infty$ (see SM). While Eqs.~\eqref{eq:kinetic}-\eqref{eq:collisionK} predict heating for $E_D>\pi/2$ {only}, heating for $E_D<\pi/2$ goes beyond this approximate description. {The condition $ E_D <\pi / 2 $ can be seen as the generalization, for Floquet disordered systems, of the fast-driving condition securing a long prethermal plateau for clean Floquet systems \cite{Abanin2015,Kuwahara2016}.}

{\it {Wave c}ondensation.--}  
{Importantly, the total quasi-energy $E_{tot}$ of the initial state is lowered by lowering the initial angle $\theta_0$ {(see SM)}.}
{Starting in the prethermal regime and decreasing $\theta_0$,} the chemical potential $\mu$ increases until it hits the lower Floquet band edge $-E_D$ for some critical value $\theta_0^c$ and a condensation process occurs. For $\theta_0 \leq  \theta_0^c$, $\mu = -E_D$, excited quasi-energy levels cannot accommodate more particles and the lowest mode gets macroscopically occupied. The condensed fraction reads $\rho_{c}=1-T\int_{-E_D}^{E_D}\frac{d\varepsilon}{2\pi} \frac{\nu(\varepsilon)}{\varepsilon+E_D}$ where $T$ is the effective temperature of the remaining prethermal part~\cite{Thermal_BEC3}.
The left panel of Fig.~\ref{fig:cond} shows the quasi-stationary $f_S(\varepsilon)$ obtained for $\theta_0=0.32$ 
and {$f_{RJ}(\varepsilon)$ obtained at $\mu=-E_D$, $T$ being the only fitting parameter. Both match at large $\varepsilon$, but $f_{RJ}(\varepsilon)$ totally fails at low $\varepsilon$.} This accumulation of population in {the lowest energy mode} signals condensation. Concomitantly, as seen in the right panel of Fig.~\ref{fig:cond}, {the quasi-stationary} $n_S(\theta)$ deviates largely from {$n_{eq}(\theta) = T\int_{-E_D}^{E_D} \frac{d\varepsilon}{2\pi} \frac{A(\varepsilon,\theta)}{\varepsilon+E_D}$} in the vicinity of $\theta=0$, which corresponds precisely to the location of the lowest energy states.   
It can be decomposed into thermal and condensed components, leading to a bimodal $\theta$-distribution reminiscent of the celebrated Bose-Einstein condensation signature~\cite{Anderson1995,Davis1995}, {$n_S(\theta) = n_c(\theta) + n_{eq}(\theta)$}. To our knowledge, {wave condensation} happening in the prethermal plateau of a Floquet system had never been observed before.

{\it Phase diagram.--}
We check 
the Onsager-Penrose criterion for condensation~\cite{Onsager_Pensore,Onsager_Pensore_GPE} by computing the time coarse-grained $1$-body density matrix given by
\begin{equation}
\langle\theta|\rho^{(1)}(t)|\theta'\rangle = {\frac{1}{\Delta t+1}} \sum_{t'=t}^{t+\Delta t} \psi^*(\theta', t') \psi(\theta, t')
\label{eq:PenOn}
\end{equation}
with $\Delta t=30$.
This coarse-graining emulates a mixed state from the pure state $\psi (\theta,t)$~\cite{Goral2002,Onsager_Pensore_GPE}. Writing $\rho^{(1)}(t) = \sum_n P_n(t) |\Phi_n(t)\rangle\langle \Phi_n(t)|$, normalization implies $\sum_n P_n(t) = 1$ and $P_n$ represents the occupation probability of the mode $|\Phi_n\rangle$. Sorting the probabilities by descending order $P_1 \geq P_2 \geq P_3 \geq \cdot\cdot\cdot$, a 'macroscopic' probability $P_1\gg P_2$ signals condensation of the system into the eigenmode $|\Phi_1\rangle$. In the left panel of Fig.~\ref{fig:phaseportrait}, we plot the population gap $\Delta P=P_1-P_2$ obtained at $t=10^4$ in the $(K,\theta_0)$ plane. We see that condensation occurs below the threshold $E_D = K+W=\pi/2$ (i.e. $K$ below the black dashed line in Fig.~\ref{fig:phaseportrait}a) where {the approximate Eqs.~\eqref{eq:kinetic}-\eqref{eq:collisionK} predict a suppression of heating}. Furthermore, as {mentioned} above, condensation {indeed occurs for $\theta_0$ below the  $\theta_0^c(K)$ critical line (red dots in Fig.~\ref{fig:phaseportrait}a). } 

Saliently, the duration of this prethermal plateau can be made exponentially large, as shown in Fig.~\ref{fig:phaseportrait}b where we plot $\tau_{boil}(K,\theta_0)$ (see SM). This key observation should allow for an observation of {RJ prethermalization and wave} condensation in experiments with Floquet systems.
Note that we have verified that the measured quantities did not show significant finite-size effects (see SM).

{\it Conclusion.--}  We have studied the nonequilibrium dynamics of a disordered Floquet system subjected to a nonlinear Gross-Pitaevskii interaction. {When} Floquet quasi-energy bands are sufficiently narrow and well separated, inter-band transitions are forbidden and heating is strongly suppressed. {This condition generalizes to disordered Floquet system the fast-driving condition for clean Floquet systems }\cite{Abanin2015,Kuwahara2016}. It allows the system to reach a prethermal plateau where it stays for an exponentially-long time before heating processes ultimately boil it to an infinite-temperature featureless state. In the prethermal plateau, low-energy physics takes place in the form of Rayleigh-Jeans prethermalization, {and wave condensation at low quasi-energies.

Our predictions are based on a variant of the kicked rotor already realized with ultracold atoms and could therefore be tested in such experiments. It was argued that GP interaction in $p$ is a good description of the spatial interaction at weak nonlinearities \cite{QKR_GPE2, PhysRevA.101.043624}. {Narrow Floquet bands can be experimentally achieved by working in the vicinity of quantum resonances \cite{izrailev1980quantum, wimberger2003quantum}. {Very recently, the many-body kicked rotor has been the subject of experimental \cite{cao2021prethermal} and theoretical \cite{vuatelet2021effective} studies on prethermalization. Our study proposes new regimes and different characterizations of this physics.}
	 The Gross-Pitaevskii equation also appears in nonlinear optics.} Experiments with disordered photonic lattices \cite{schwartz2007transport, PhysRevLett.100.013906, levi2012hyper} could also address the prethermalization properties we have {discussed}. {Future work could address} superfluid or turbulent transport in such Floquet systems \cite{PhysRevA.82.011602, PhysRevA.80.033615}.

\begin{acknowledgments}
	We thank M.~Albert, J.~Billy, N.~Cherroret, D.~Delande, O.~Giraud, D.~Gu\'ery-Odelin, N.~Mac\'e and D. Ullmo for interesting discussions.
	This study has been supported by the French National Research Agency  (ANR) under  projects COCOA ANR-17-CE30-0024, MANYLOK  ANR-18-CE30-0017 and GLADYS ANR-19-CE30-0013, the   EUR  grant NanoX  No. ANR-17-EURE-0009  in  the  framework  of the ``Programme des Investissements d'Avenir'', and by the Singapore Ministry of Education Academic Research Fund Tier I (WBS No. R-144-000-437-114). Computational resources were provided by the facilities of Calcul en Midi-Pyr\'en\'ees (CALMIP) and the National Supercomputing Centre (NSCC), Singapore.
\end{acknowledgments}

%

\pagebreak
\widetext
\begin{center}
	\textbf{\large Supplementary Material for \\"Prethermalization and wave condensation in a nonlinear disordered Floquet system"}
\end{center}
\setcounter{equation}{0}
\setcounter{figure}{0}
\setcounter{table}{0}
\setcounter{page}{1}
\makeatletter
\renewcommand{\theequation}{S\arabic{equation}}
\renewcommand{\thefigure}{S\arabic{figure}}
\renewcommand{\bibnumfmt}[1]{[S#1]}
\renewcommand{\citenumfont}[1]{S#1}

\section{Floquet-Boltzmann kinetic equation}
{We give here more details about the derivation of the kinetic equation (5)-(6) for the Floquet system considered. Very generally, the time-periodic Floquet modes $\phi_\alpha(p,t)$ of the linear quantum kicked rotor satisfy
	\begin{equation}
	\phi_\alpha(p,t) = \phi_\alpha(p, t+1) \hspace{1.5cm}   [\mathcal{H}(t) - \mathrm{i} \partial_t] \phi_\alpha(p,t) = \varepsilon_\alpha \, \phi_\alpha(p,t)  \hspace{1.5cm}  \sum_{p\in\mathbbm{Z}} \phi_\alpha^*(p,t) \phi_\beta(p,t) = \delta_{\alpha\beta}
	\end{equation}
	with a continuum of quasi-energies $\varepsilon_\alpha \in [-\pi,\pi]$. In our numerical simulations, we have however considered a finite-size momentum basis set restricted to $N$ states, {$-N/2 \leq p \leq N/2-1$}, with periodic boundary conditions. The Hilbert space of the system having dimension $N$, we also need to consider $N$ linear Floquet modes $\phi_\alpha(p,t)$ ($1\leq \alpha \leq N$) and expand $\psi(p,t)$ over these $N$ Floquet modes
	\begin{equation}
	\psi(p,t)= \frac{1}{\sqrt{N}} \,\sum_{\alpha=1}^N c_\alpha(t) \, \phi_\alpha(p,t),
	\end{equation}
	with the updated nomalisation conditions {
		\begin{equation}
		\sum_{p=-N/2}^{N/2-1} \phi_\alpha^*(p,t) \phi_\beta(p,t) = \delta_{\alpha\beta} \hspace{1.5cm} \sum_{p=-N/2}^{N/2-1} |\psi_(p,t)|^2= 1.
		\end{equation}}
	In turn, these two conditions imply the normalisation condition $\sum_{\alpha=1}^N |c_\alpha|^2 = N$.}

From the nonlinear Schr$\ddot{\rm o}$dinger equation~(1) of the manuscript, it is easy to see that the modal coefficients $c_{\alpha}$ satisfy the following equation of motion
\begin{equation}
\mathrm{i} \frac{dc_{\alpha}}{dt} = \varepsilon_{\alpha}c_{\alpha} + g \, \sum_{\beta\gamma\delta} W_{\alpha\beta\gamma\delta} \, c^{*}_{\beta}c_{\gamma}c_{\delta},
\label{Eq2}
\end{equation}
featuring the 4-point Floquet modes correlator
\begin{equation}
W_{\alpha\beta\gamma\delta}(t)= \frac{1}{N} \, \sum_{p} \phi^{*}_{\alpha}(p,t)\phi^{*}_{\beta}(p,t) \phi_{\gamma}(p,t) \phi_{\delta}(p,t).
\label{Eq3}
\end{equation}
At this stage, Eqs.\eqref{Eq2}-\eqref{Eq3} are exactly Eqs.(148-149) in \cite{WT_theory2} and Eqs.(6-7) in \cite{WT_theory3} with the salient difference that our correlator is time-dependent and, most importantly, time-periodic $W_{\alpha\beta\gamma\delta}(t+1) = W_{\alpha\beta\gamma\delta}(t)$ since the Floquet modes themselves are time-periodic. We thus expand $W$ in Fourier series:
\begin{equation}
W_{\alpha\beta\gamma\delta}(t)=\sum_{m\in\mathbbm{Z}} \, W_{\alpha\beta\gamma\delta}^{(m)} \, e^{-2\mathrm{i}\pi m t}.
\label{Eq4}
\end{equation}
Now, changing variables $c_\alpha \to \tilde{c}_\alpha = c_\alpha \, \exp(\mathrm{i}\varepsilon_\alpha t)$ and dropping the tilde to ease the notation, Eq.\eqref{Eq2} rewrites
\begin{equation}
\mathrm{i} \frac{dc_{\alpha}}{dt} = g \, \sum_{m\beta\gamma\delta} e^{\mathrm{i} (\varepsilon_\alpha+\varepsilon_\beta - \varepsilon_\gamma - \varepsilon_\delta +2\pi m)t} \, W_{\alpha\beta\gamma\delta}^{(m)} \, c^{*}_{\beta}c_{\gamma}c_{\delta}. 
\label{Eq5}
\end{equation}
At this point, one can follow the usual steps of the derivation of the irreversible kinetic equation by using the random phase approximation developed in \cite{Zakharov2004} and by taking the continuum limit (large $N$ assumption) as detailed in pages 51-57 and Appendix A7 of \cite{WT_theory2} and in \cite{WT_theory3}. By doing so in our case, the discrete sums become integrals over quasi-energies between $-\pi$ and $\pi$ and introduce products of the disorder-averaged density of states per unit volume $\nu(\epsilon) = \frac{1}{N} \, \overline{\sum_\alpha \delta (\varepsilon-\varepsilon_\alpha)}$. After tedious calculations, one gets Eqs.(4-5) in the main text with $R^{(m)} = \overline{|W^{(m)}|^2}$. Do note that the large-$N$ limits of sums are obtained through:
\begin{equation}
\frac{1}{N} \, \sum_\theta (\cdot\cdot\cdot) \to \int_{-\pi}^{\pi} \frac{d\theta}{2\pi} \, (\cdot\cdot\cdot) \hspace {0.75cm} \frac{1}{N} \, \sum_\alpha (\cdot\cdot\cdot) \to \int_{-\pi}^{\pi} d\varepsilon \, \nu(\varepsilon) \, (\cdot\cdot\cdot)
\end{equation}

A remark is in order here. In the usual wave turbulence (WT) approach {\cite{WT_theory2}}, the free Hamiltonian does not generate any randomness in the evolution of the system: One starts with an initial state with static random phases and uses averages over these initial random phases to derive the kinetic equation. Here, we instead start with a perfectly phase-coherent initial state and it is the subsequent dynamics generated by the disordered Floquet evolution operator that brings in phase randomness. In this case, as argued in \cite{Thermal_BEC3} for weakly nonlinear spatially disordered systems, one has just to replace the averages over the static initial random phases in the WT approach by disorder configuration averages {(the number of disorder configurations used in our numerical simulations is denoted by $N_d$)}. This prescription is at least valid as long as the Boltzmann time $\tau_B$
associated to disorder is much smaller that the equilibration time $\tau_{eq}$ associated to the nonlinear couplings. In other words, disorder quickly isotropizes the system before nonlinearity starts to significantly modify the energy distribution. 

It is crucial to note that the Floquet kinetic equation does conserve particle number but does not conserve necessarily the total initial quasi-energy. Indeed, the $\delta(\varepsilon_\alpha+\varepsilon_\beta - \varepsilon_\gamma - \varepsilon_\delta +2\pi m)$ term coming along with each collision kernel $R^{(m)}$ shows that Floquet systems exhibit collision processes satisfying energy conservation up to nonzero multiples of the driving frequency $\Omega=2\pi$ (in Eq.~(1) the kicking period is $T=1$) in addition to the strictly resonant energy transfer term $m=0$. For periodically-driven systems, these $m\neq 0$ processes are the equivalent of Umklapp scattering processes in space-periodic systems, where "colliding" wavevectors are folded back to the first Brillouin zone through appropriate reciprocal lattice vector translations. Note that these Umklapp processes heat and boil  the system. In this approximate description, it is only when these $m\neq 0$ processes are forbidden (i.e. when $E_D < \pi/2$) that the total energy of the system is conserved, in which case the system can reach and stay in the Rayleigh-Jeans (RJ) thermal equilibrium and, if appropriate conditions are met, undergo the classical wave condensation scenario (see below). However, the heating observed at long times $t\gg \tau_{boil}$ for $E_D<\pi/2$ goes beyond this Floquet kinetic description. 

\section{Scaling of the CBS and CFS decay time and equilibration time with interaction strength}

\begin{figure}
	\centering
	\includegraphics[width=0.8\linewidth]{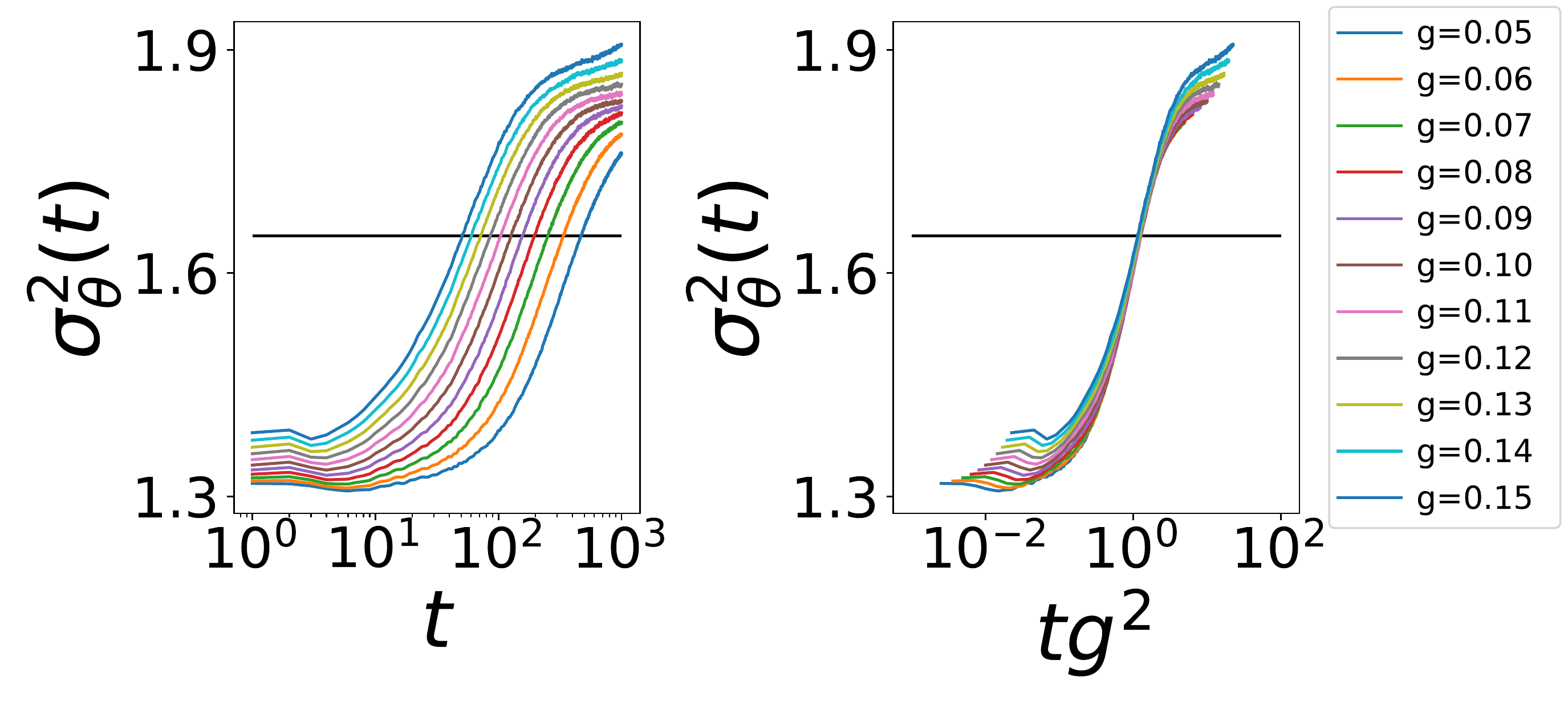}
	\caption{$\theta$-variance $\sigma^2_\theta(t)$ as a function of time for different values of the Gross-Pitaevskii strength $g$. Left panel: $\sigma^2_\theta(t)$ is plotted against $t$ for different $g$. Right panel: the same $\sigma^2_\theta(t)$ is now plotted against $tg^2$ for different $g$. All curves collapse in the intermediate region. The horizontal black line shows the threshold value $\sigma^2_\theta = 1.65$ used to extract $\tau_{eq}$ for different $g$ (see text). {The parameters are $K=1$, $W=0.4$, $\theta_0=1.09$ $N=1024$ and $N_d=500$.}
	}
	\label{fig:tau_eq_t_g}
\end{figure}

\begin{figure}
	\centering
	\includegraphics[width=0.8\linewidth]{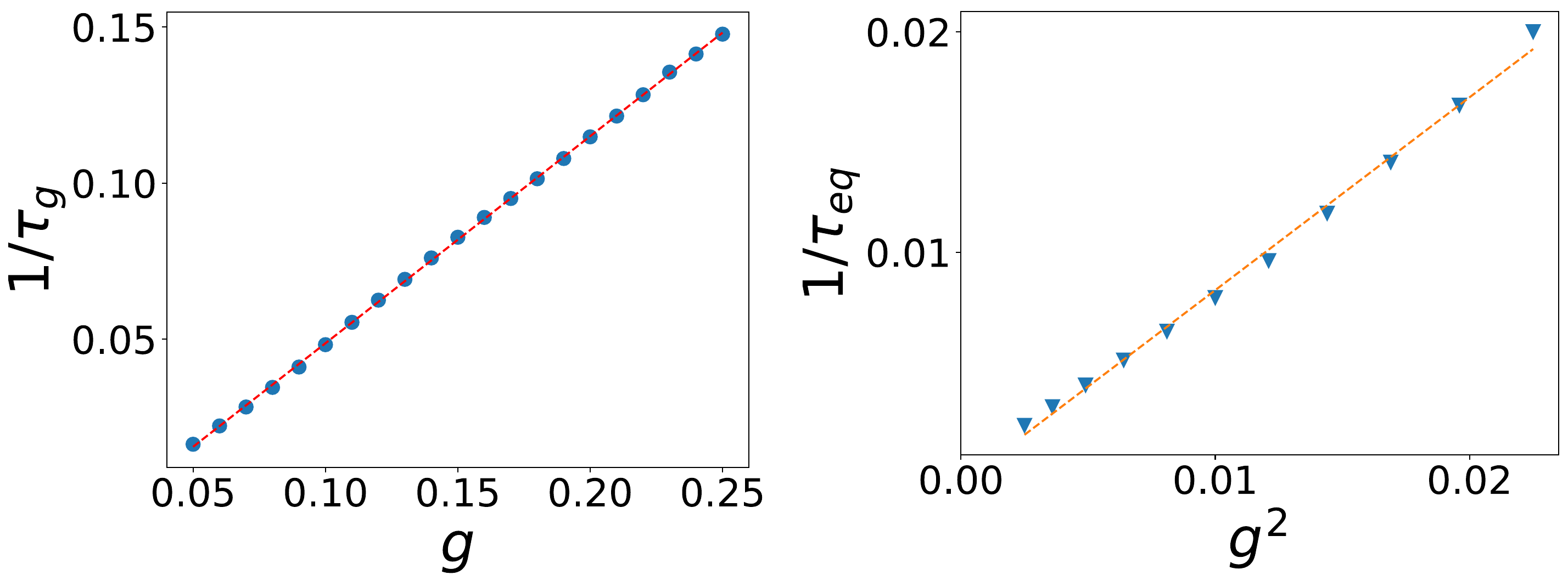}
	\caption{{Dependence of the characteristic CBS and CFS decay time $\tau_{g}$ and of the equilibration time $\tau_{eq}$ on the Gross-Pitaevskii strength $g$. Left panel: plot of $\tau^{-1}_{g}$ versus $g$ for $K=1$, $W=0.4$, $\theta_0=1.57$, $N=1024$ and $N_d=500$. $\tau_{g}$ is extracted from an exponential fit of the time decay of the CBS and CFS peaks. The red dashed line is a linear fit to the data. Right panel: Equilibration time $\tau_{eq}$ versus $g^2$ for $K=1$, $W=0.4$, $\theta_0=1.09$ $N=1024$ and $N_d=500$. $\tau_{eq}$ is defined by the time at which the $\theta$-variance $\sigma^2_\theta(t)$ reaches the threshold value $\sigma^2_\theta = 1.65$. The orange dashed line is a linear fit to the data.}
	}
	\label{fig:tau_vs_g}
\end{figure}


As seen in Fig.~1 of the main text, when nonlinearity is switched on, the CBS and CFS peaks quickly die off over some characteristic decay time $\tau_{g}$ for the chosen system parameters. Meanwhile, the nonlinearity term also redistributes the energy over different Floquet modes and the system reaches either the RJ distribution or wave condensation after a larger time scale $\tau_{eq}$. Then the system may stay at the prethermal plateau until $\tau_{\textrm{boil}}$.  For our system with an interaction term local in $p$, it was argued that the nonlinear time scale $\tau_{g}$ corresponds to the interaction energy stored in a localization volume, hence $\tau_{g} \propto g^{-1}$ \cite{Cherroret2014} (see however \cite{PhysRevResearch.2.033349} where $\tau_{g} \propto g^{-2}$ is predicted at very small $g$), and the equilibration time $\tau_{\textrm{eq}}\propto g^{-2}$ \cite{Thermal_BEC3}. 

{In Fig.~\ref{fig:tau_eq_t_g} left panel, we characterize $\tau_{eq}$ by the time at which the $\theta$-variance $\sigma^2_\theta(t)$ reaches a given threshold value (indicated by the horizontal black line in Fig.~\ref{fig:tau_eq_t_g}) between its initial value and its quasi-stationary value in the prethermal plateau. On the other hand, we define} $\tau_{g}$ by fitting the early time dynamics of the CBS and CFS peaks by an exponential decay
\begin{equation}
n(\pm \theta_0,t) \sim n(\pm \theta_0,t=0) \exp(-t/\tau_{g}).
\end{equation}
In Fig.~\ref{fig:tau_vs_g}, we have plotted the extracted inverse decay rates $\tau^{-1}_{g}$ and $\tau_{\textrm{eq}}^{-1}$ as a function of $g$ and $g^2$. As one can see, the agreement with the conjectured scaling relations, $\tau_{g} \propto g^{-1}$ and $\tau_{\textrm{eq}}\propto g^{-2}$, is pretty good. Note however that the data presented in this paper correspond to values of the nonlinear interaction strength significantly larger than those considered in \cite{PhysRevResearch.2.033349}. Data (not shown) for $\tau_g$ with $g \in [10^{-3},10^{-2}]$ seem compatible with the prediction $\tau_{g} \propto g^{-2}$ of \cite{PhysRevResearch.2.033349}. {On the other hand, Fig.~\ref{fig:tau_eq_t_g} right panel shows that the curves of $\sigma^2_\theta(t)$ for different values of $g$ all collapse onto each other in the transition regime between the initial stage and the prethermal plateau when plotted as a function of $t g^2$. This again validates $\tau_{\textrm{eq}}\propto g^{-2}$.}


\section{Variation of the total quasi-energy with initial position}

\begin{figure}
	\centering
	\includegraphics[width=1\textwidth]{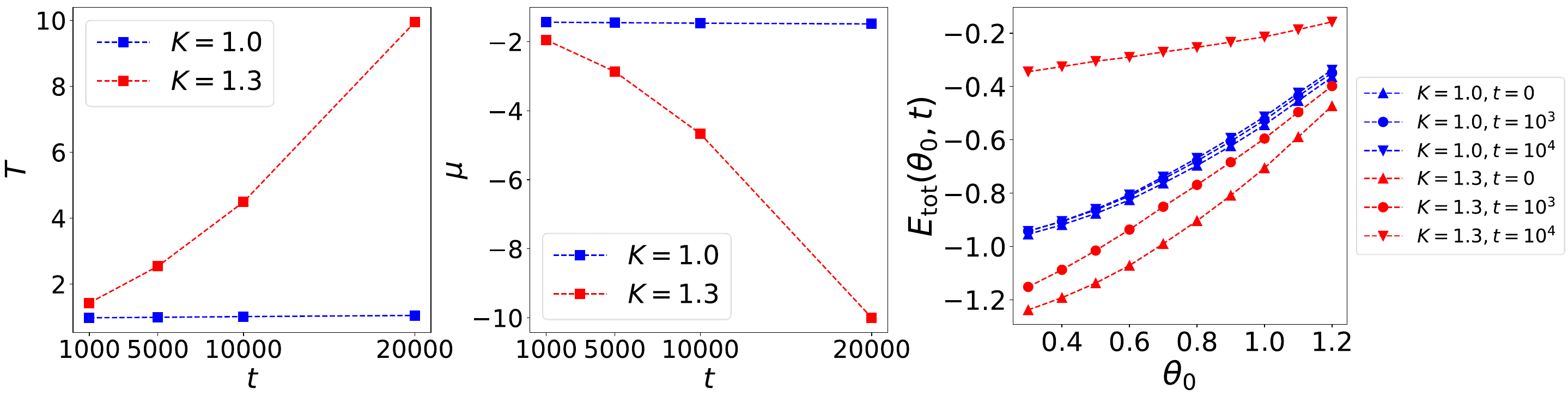}
	\caption{{Time dependence of the temperature $T$ (left panel) and chemical potential $\mu$ (middle panel) obtained for the initial state $\theta_0=1.05$ at $K=1$ (blue) and $K=1.3$ (red). Both $T$ and $\mu$ are extracted by fitting $f_\varepsilon(t)$ with the Rayleigh-Jeans distribution. Right panel: Graph of the total quasi-energy per mode $E_{tot}$ as a function of $\theta_0$ at 3 different times $t$ for the same two $K$ values. For all panels, the other system parameters are fixed at $g=0.1$, $W=0.4$, $N=1024$ and $N_d=1000$ disorder configurations. With $W=0.4$, the Umklapp processes $m \neq 0$ are active for $K=1.3$ ($E_D=K+W > \pi/2$) and suppressed for $K=1.0$ ($E_D=K+W < \pi/2$). As one can see, there is almost no heating in the absence of Umklapp processes: Both $T$, $\mu$ and $E_{tot}$ stay essentially the same in the considered time range $t\in[0,10^4]$. The situation is markedly different when Umklapp processes are present: Both $T$, $\mu$ and $E_{tot}$ change significantly.} 
	}
	\label{fig:e_vs_x0}
\end{figure}

The linear contribution to the total quasi-energy per mode of the system, $E_{tot}=\int_{-\pi}^{\pi} d\varepsilon \, \varepsilon \, \nu(\varepsilon) \, f_\varepsilon(t)$, see \cite{Thermal_BEC2}, depends on the initial rotor angle $\theta_0$ through the initial state {$\psi(p,t=0)= \frac{1}{\sqrt{N}} \exp(-\mathrm{i} p \theta_0)$}.  {The other system parameters being fixed,} low values of {$\theta_0$} correspond to low quasi-energies $E_{tot}$. {Indeed, for our system, $\theta$ corresponds to wave-vector $k$ in spatially disordered systems}. It is worth mentioning that $E_{tot}$ is symmetric in $\theta_0$ because the kicked rotor Hamiltonian is symmetric in $\theta$.

{In the right panel of Fig.~\ref{fig:e_vs_x0}, we plot $E_{tot}$ as a function of $\theta_0$ at 3 different times ($t=0$, $t=10^3$ and $t=10^4$) and for 2 different kick strengths ($K=1$ and $K=1.3$). The interaction and disorder strengths are fixed at $g=0.1,W=0.4$. For $K=1.3$, the Umklapp processes $m\neq 0$ are present ($E_D=K+W \geq \pi/2$) and $E_{tot}$ increases fast with time. On the other hand, the Umklapp processes $m\neq 0$ are suppressed for $K=1$ ($E_D=K+W < \pi/2$) and $E_{tot}$ is almost conserved and independent of time in the range $t\in[0,10^4]$. Note that $E_{tot}$ at $t=0$ reads
	\begin{equation}
	E_{tot}(\theta_0,t=0) = \int_{-\pi}^{\pi} \rm d\varepsilon \,  \varepsilon \, A_\varepsilon(\theta_0)
	\end{equation}
	since $c_{\alpha}(t=0) = \phi^*_\alpha(\theta_0, t=0)$ and $ \nu(\varepsilon)f_{\varepsilon}(t=0) = \frac{1}{N} \overline{\sum_{\alpha=1}^{N} \delta(\varepsilon - \varepsilon_\alpha) \vert c_\alpha(t=0)\vert^2} = \frac{1}{N} \overline{\sum_{\alpha=1}^{N} \delta(\varepsilon - \varepsilon_\alpha) \vert \phi_\alpha(\theta_0, t=0)\vert^2}$ is the spectral function $A_\varepsilon(\theta_0)$.}

Moreover, we observe that the quasi-energy distribution $f_\varepsilon(t)$ can be fitted by a Rayleigh-Jeans distribution not only in the prethermal regime but also in the heating regime $t\gtrsim \tau_{boil}$. {In the left and middle panels of Fig.~\ref{fig:e_vs_x0},} we show the extracted temperature $T(t)$ and chemical potential $\mu(t)$ at different times {for the same values of $K$ as in the right panel}. Both $T$ and $\mu$ stay essentially constant if the system resides in the prethermal plateau $\tau_{eq} \ll t \ll \tau_{boil}$ (case $K=1$ in Fig.~\ref{fig:e_vs_x0} where $E_D = K+W < \pi/2$). On the contrary,  in the heating regime ($K=1.3$ in Fig.~\ref{fig:e_vs_x0}) $T(t) \rightarrow +\infty$ and $\mu(t)\rightarrow -\infty$, while $T(t)/\mu(t) \rightarrow -1$ when $t \rightarrow \infty$. In other words, the system reaches an infinite temperature state with flat distributions $f_\varepsilon (t)$ and $n(\theta,t)$ when $t\gg \tau_{boil}$.

\section{Determination of $\tau_{boil}$ from the dynamics of the spatial distribution}

As discussed in the manuscript, after a time scale $\tau_{eq}$, the system reaches a quasi-stationary state well described by a Rayleigh-Jeans thermal distribution or showing signatures of wave condensation, up to a characteristic time $\tau_{boil}$ after which the system heats up rapidly. Here we demonstrate that the duration of the prethermal plateau $(\tau_{boil}-\tau_{eq})$ scales exponentially with the parameter $K$. {To extract $\tau_{boil}$}, we plot $\sigma^2_\theta(t)$ and its logarithmic derivative as a function of $\log_{10}t$, see Figure~\ref{fig:sigma_vs_t}. There are two clear peaks in the logarithmic derivative curve: {The first one corresponds to $\tau_{eq}$ (see Fig.~\ref{fig:tau_eq_t_g} for a more precise numerical determination)} while the second one gives $\tau_{boil}$. One can then study the dependence of $(\tau_{boil}-\tau_{eq})$ on the different parameters of the system, in particular $K$. 

{The upper panels of Fig.~\ref{fig:t_boil} show that all curves $\sigma^2_\theta(t)$ obtained for different values of $K$ almost collapse onto each other up to a timescale $\sim 10^2$. This agrees with the expectation that $\tau_{eq}\sim g^{-2}$ ($g=0.1$ in the plot) does not strongly depend on $K$. The lower panels of Fig.~\ref{fig:t_boil} show $\tau_{boil}$ as a function of $K$. For $\theta_0=1.09$ (lower left panel),} we observe that $\tau_{boil}$ scales exponentially with $K$, validating a very long prethermal plateau in our system $(\tau_{boil}-\tau_{eq})\approx \tau_{boil}$ when $E_D = K + W \lesssim \pi/2$. {Moreover, when $\theta_0$ is sufficiently small (as shown in the lower right panel where $\theta_0=0.01$), $\tau_{boil}$ decreases an order of magnitude faster with $K$ after the heating threshold $E_D=\pi/2$ given by the Floquet kinetic equation, Eqs (5)-(6) in the main text, is crossed.} This is a clear signature of the heating effect generated by the Umklapp terms $m \neq 0$.

{The prethermal properties that we observe in our disordered Floquet system can be seen as a generalization of those of clean Floquet systems in the fast-driving regime \cite{Abanin2015,Kuwahara2016}.} In fact, changing the driving frequency would amount in our case to multiplying $ W $ and $ g $ by a certain factor, which is also equivalent to changing the value of $ K $.


\begin{figure}[h!]
	\centering
	\includegraphics[width=0.6\textwidth]{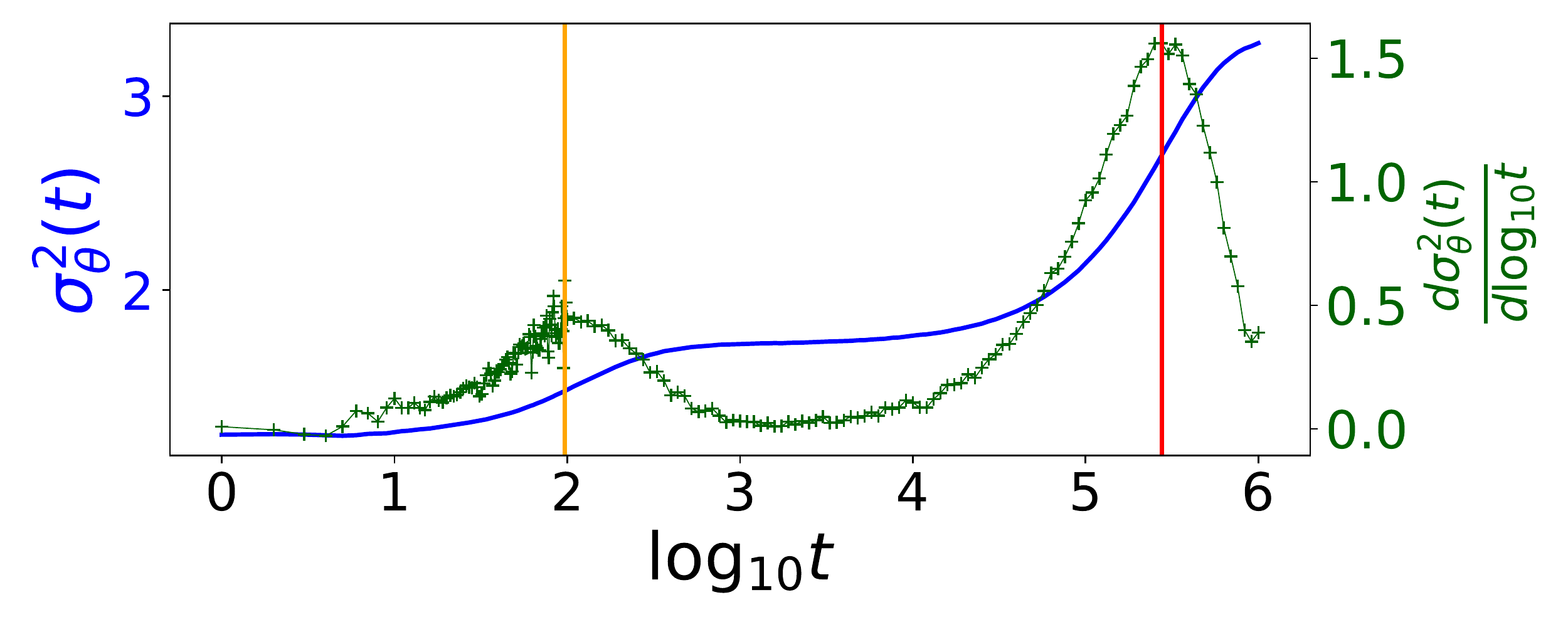}
	\caption{{$\theta$-variance $\sigma^2_\theta(t)$ (blue line) and its logarithmic derivative (green crosses connected by a green line to guide the eye) versus $\log_{10}t$. The timescales $\tau_{eq}$ and $\tau_{boil}$, marked by the orange and red vertical lines, are given by the local maxima of the logarithmic derivative {(see Fig.~\ref{fig:tau_eq_t_g} for a more precise numerical determination of $\tau_{eq}$)}. The system parameters are $K=1.0$, $\theta_0=1.05$, $g=0.1$, $W=0.4$, $N=1024$ and $N_d=1200$ disorder configurations. The duration of the prethermal plateau is given by  $(\tau_{boil}-\tau_{eq})$.}
	}
	\label{fig:sigma_vs_t}
\end{figure}

\begin{figure}
	\centering
	\includegraphics[width=0.9\textwidth]{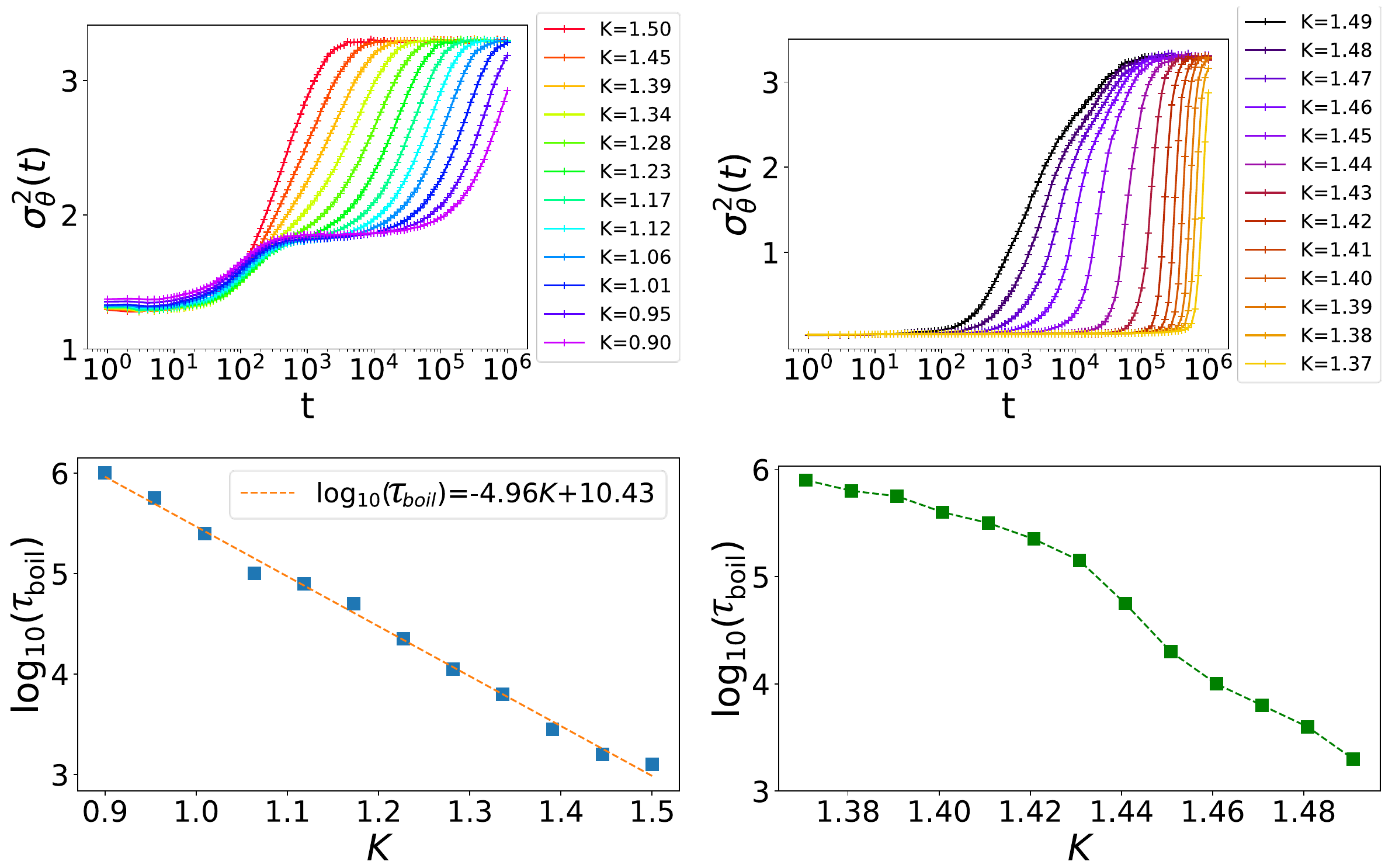}
	\caption{{Upper panels: $\theta$-variance $\sigma^2_\theta(t)$ as a function of time for different values of $K$. Lower panels: Log-linear plot of $\tau_{boil}$ versus $K$. The parameters are $\theta_0=1.09$, $g=0.1$, $W=0.4$ and $N_d=250$ in the two left panels while $\theta_0=0.01$, $g=0.05$, $W=0.1$ and $N_d=50$ in the two right panels. In all panels, $N=1024$. The linear fit in the lower left panel (orange dashed line) shows that $ \tau_{boil}$ scales exponentially with K. As further seen in the lower right panel, two decay regimes happen at smaller initial angles: $ \tau_{boil}$ decreases faster with $K$ when $K>1.44$. This kink value is close to the heating threshold $K =\pi/2-W \approx 1.47$ predicted by the Floquet kinetic description.}} 
	\label{fig:t_boil}
\end{figure}

\section{System size dependence of the Onsager-Penrose criterion for condensation and of $\tau_{boil}$}

{In the main text, we have checked the Onsager-Penrose criterion for condensation~\cite{Onsager_Pensore,Onsager_Pensore_GPE} by computing and diagonalizing the time coarse-grained $1$-body density matrix $\langle\theta|\rho^{(1)}(t)|\theta'\rangle$, see Eq.~(7), for $N=1024$. Condensation is signalled by the existence of a "macroscopic" population gap between the largest and second largest eigenvalues of $\rho^{(1)}(t)$. Here, we consider this population gap $\Delta P=(P_1-P_2)$, obtained at $t=10^4$, for different system sizes $N$, see Fig.~\ref{fig:rhoc_N}, left panel. We see} that $\Delta P$ remains almost constant as the system size $N$ increases, at least up to $N=10^5$. This suggests that wave condensation survives in the thermodynamic limit $N \rightarrow \infty$.

\begin{figure}
	\centering
	\includegraphics[width=0.45\textwidth]{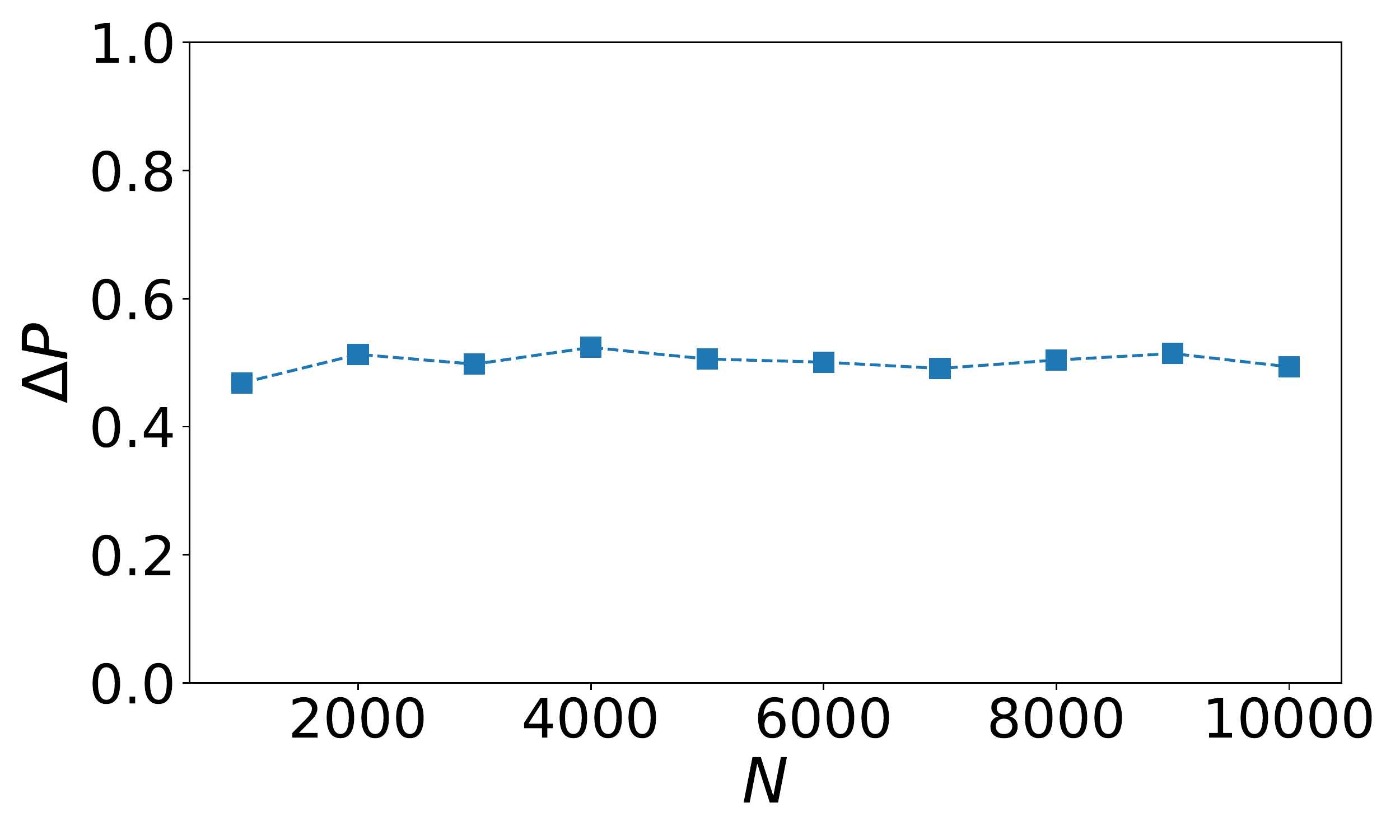}
	\includegraphics[width=0.4\textwidth]{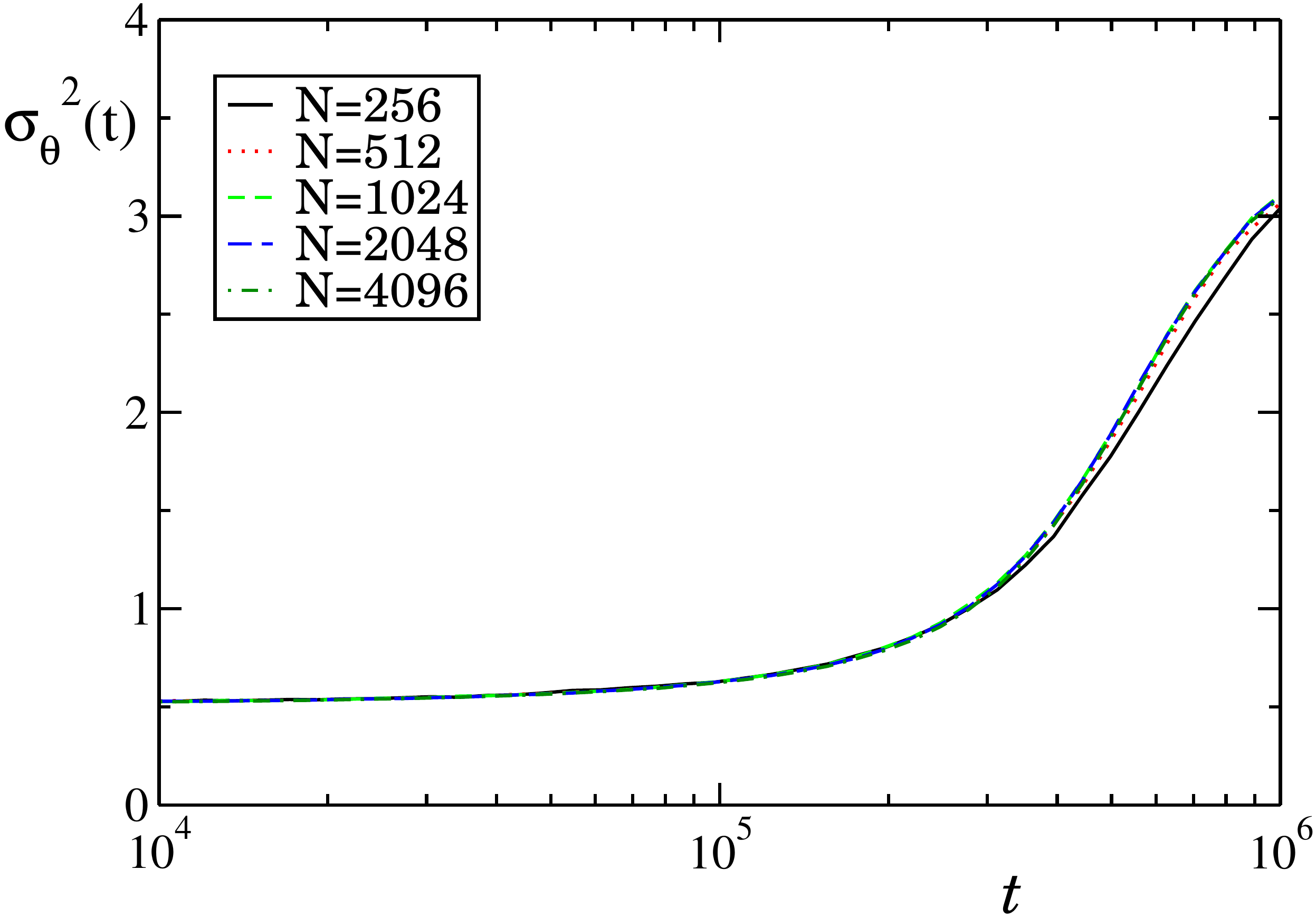}
	\caption{Left panel: Population gap $\Delta P$ of the time coarse-grained $1$-body density matrix, obtained at $t=10^4$ for different system sizes $N$.  {$\Delta P$ stays constant up to $N=10^5$ suggesting that wave condensation is robust in the thermodynamic limit $N \rightarrow \infty$.} Right panel: $\theta$-variance $\sigma^2_\theta(t)$ as a function of time for different values of $N$. The curves for different $N\gtrsim 1024$ collapse onto each other, implying that $\tau_{boil}$ does not suffer from significant system size effects. The system parameters are $K=1$, $W=0.4$, $\theta_0=0.32$ and $g=0.1$.
	}
	\label{fig:rhoc_N}
\end{figure}

We have also studied the system size dependence of $\tau_{boil}$. As shown in the right panel of Fig.~\ref{fig:rhoc_N}, the curves for the $\theta$-variance $\sigma^2_\theta(t)$ as a function of time for increasing system size collapse onto each other for $N\gtrsim 1024$ which indicates that $\tau_{boil}$ does not suffer from strong finite size effects.

\end{document}